\documentclass[twocolumn,showpacs,preprintnumbers,amsmath,amssymb]{revtex4}
\usepackage{graphicx,color}

\begin{document}

\title{Flexoelectricity and competition of time scales in electroconvection}

\author{Tibor T\'oth-Katona, N\'{a}ndor \'{E}ber,
\'{A}gnes Buka} \affiliation{Research Institute for Solid State
Physics and Optics, Hungarian Academy of Sciences, H-1525 Budapest,
P.O.B. 49, Hungary}
\author{Alexei Krekhov}
\affiliation{Physikalisches Institut, Universit\"at Bayreuth,
D-95440 Bayreuth, Germany}
\date{\today}

\begin{abstract}
Novel behavior in electroconvection (EC) has been detected in
nematic liquid crystals (NLCs) under the condition of comparable
timescales of the director relaxation and the period of the
driving ac voltage. The studied NLCs exhibit standard EC (s-EC) at
the onset of the instability, except one compound in which
nonstandard EC (ns-EC) has been detected. In the relevant
frequency region, the threshold voltage for conductive s-EC bends
down considerably, while for dielectric s-EC it bends up strongly
with the decrease of the driving frequency. We show that inclusion
of the flexoelectric effect into the theoretical description of
conductive s-EC leads to quantitative agreement, while for
dielectric s-EC a qualitative agreement is achieved. The frequency
dependence of the threshold voltage for ns-EC strongly resembles
that of the dielectric s-EC.
\end{abstract}
\maketitle

\section{\label{sec:intro}Introduction}

Electroconvection (EC) in nematic liquid crystals (NLCs) is an
extensively studied example of electric field induced instabilities
\cite{kramer,kramer1,Buka05}. It also serves as a convenient model
system for investigating pattern forming phenomena in complex,
anisotropic fluids driven out of equilibrium.

NLCs have a uniaxial orientational order which is represented by the
unit vector $\mathbf{n}$, the director. Consequently, NLCs possess
direction dependent physical properties. The dielectric anisotropy
$\varepsilon_a=\varepsilon_{\|}-\varepsilon_{\bot}$ and the
conductivity anisotropy $\sigma_a=\sigma_{\|}-\sigma_{\bot}$ are two
key parameters in EC ( ${\|}$ denotes the value along, and ${\bot}$
perpendicular to the director). While $\varepsilon_a$ governs the
electric torque acting on the director in the presence of electric
fields, the director distortions lead to an elastic restoring torque
exerted on $\mathbf{n}$. Furthermore, the uniaxial symmetry of NLCs
allows for a coupling between director orientation and flow,
characterized by the 5 independent viscosity coefficients, resulting
in a viscous torque if velocity gradients and/or temporal variation
of $\mathbf{n}$ are present.

Electroconvection has most commonly been observed in planarly
oriented ($\mathbf{n}$ parallel to the substrates) layers of
nematics with $\varepsilon_a<0$ and $\sigma_a>0$, though it  shows
up at some other combinations of $\varepsilon_a$, $\sigma_a$ and
initial director orientation too \cite{Buka05}. In experiments
usually samples of $10\mu$m$\lesssim d \lesssim 100 \mu$m thickness
have been used. EC appears at a threshold rms value $U_{\rm c}$ of
the applied ac voltage $U$ of frequency $f$ as a pattern consisting
of dark and bright stripes with wavevector $\mathbf{q}$ when viewed
in a microscope. EC patterns have great morphological richness at
the onset: normal rolls ($\mathbf{q}$ parallel with the initial
$\mathbf{n}$) and oblique rolls ($\mathbf{q}$ encloses a finite
angle with $\mathbf{n}$), stationary as well as traveling patterns
have been detected depending on material parameters and on the
driving frequency.

Detailed theoretical studies in the past decades have established a
firm understanding of this pattern forming phenomenon, combining the
equations of nematohydrodynamics (anisotropic Navier-Stokes equation
for the flow plus the balance of torques acting on the director)
with those of electrodynamics assuming that the NLC has an ohmic
electrical conductivity. First it has  been shown using a simple
one-dimensional approximation that the EC pattern is excited by the
Carr-Helfrich feedback mechanism \cite{Carr,Helfrich}: spatial
director modulations induce a separation of space charges due to the
anisotropic conductivity; electrostatic forces induce a flow in form
of vortices; flow exerts a destabilizing viscous torque on the
director against restoring elastic and electric torques. When this
feedback becomes positive (at $U\geq U_{\rm c}$) a specific director
fluctuation grows up; i.e., a periodic director modulation,
characterized by a finite tilt angle with respect to the initial
alignment, develops. As the director is the optical axis of the NLC,
the tilt leads to a modulation of the refractive indices. The
pattern becomes optically detectable either with or without a
polarizer due to light focusing/defocusing effects (shadowgraph
image \cite{Rasenat89}) as well as via birefringence if crossed
polarizers are used. The complex physical phenomena involved in the
mechanism occur on three different time scales, characterized by the
director relaxation time $\tau _{\rm d} = \frac{\gamma _1
d^2}{K_{11} \pi ^2}$, the charge relaxation time $\tau _{\rm q} =
\frac{ \epsilon _0 \epsilon _{\perp}}{\sigma _{\perp}}$ and the
viscous relaxation time $\tau_{\rm v}= \frac{\rho d^2}{\alpha_4/2}$.
Here $K_{11}$ is the splay elastic modulus, $\rho$ is the density,
$\gamma _1$ is the rotational, $\alpha_4/2$ the isotropic viscosity.
Typically $\tau _{\rm d} \gg \tau _{\rm q} \gg \tau_{\rm v}$
\cite{kramer}. For example, in a $d=20\mu$m sample of nematic Phase
5 (Merck \& Co., Inc.) with a typical conductivity of
$\sigma_{\perp}=8.2 \times 10^{-8} (\Omega {\rm m})^{-1}$ at
$T=30^{\circ}$C, one has $\tau_{\rm d} = 0.46$s, $\tau_{\rm q} = 5.6
\times 10^{-4}$s and $\tau_{\rm v} = 1.4 \times 10^{-5}$s.

Though the 1-d theoretical description managed to capture the
essence of the instability mechanism, it represents a substantial
simplification of the problem, therefore the applicability of the
resulting (approximate) analytical formulas are limited. This has
necessitated later the development of a comprehensive 3-d theory,
composed of six coupled partial differential equations (PDEs), known
today as the standard model (SM) of EC \cite{Bodenschatz88}. The
equations can be conveniently made non-dimensional by measuring
lengths in units of $d/ \pi$ and time in units of $\tau_{\rm d}$.
Then via a linear stability analysis the SM can provide the
frequency dependence of the threshold voltage $U_{\rm c}(f)$, that
of the critical wave vector, as well as the spatiotemporal
dependence of the director field, the velocity field and the charge
distribution at onset. According to the model, a finite threshold of
the primary instability in planar geometry requires a positive
$\sigma_a$ and a negative (or slightly positive) $\varepsilon_a$.
Under these conditions, depending on the driving frequency $f$, two
different solution types exist. Below the crossover frequency
$f_{\rm c}$, in the conductive regime, the charge distribution
oscillates with $f$, while the director field is stationary in
leading order. In this regime, SM gives two solutions with different
spatiotemporal symmetry (mode I and mode II), however, for the
threshold behavior only the solution with the lower $U_{\rm c}$
(mode I) is relevant \cite{Dressel,Zimmer91}. Above $f_{\rm c}$, in
the dielectric regime, the situation is reversed: the charge
distribution is stationary in leading order, while the director
oscillates with $f$. Here, the model also yields two solutions (mode
III and mode IV), but the threshold behavior is described by mode
III which has lower $U_{\rm c}$ compared to mode IV. These
electroconvection scenarios, that are captured by the SM, will be
referred to as standard electroconvection (s-EC). Various
predictions of the SM have been compared with experimental results
obtained on a number of NLCs and usually an excellent agreement has
been found -- see, e.g., \cite{kramer1,Rasenat90,Schneider92}.

In s-EC, the conductive and the dielectric regimes correspond to two
competing pattern forming modes which have different $U_{\rm c}(f)$
and $U_{\rm c}(d)$ characteristics. The frequency dependence of
$U_{\rm c}$ and $q_{\rm c}$ has a concave (diverging) shape in the
conductive regime whereas they have a square root-like behavior in
the dielectric one. As far as the thickness dependence is concerned,
an inspection of the non-dimensional equations of the SM shows that
by scaling only 5 of the 6 PDEs became invariant with respect to
$d$; the equation for the potential still has an explicit $d$
dependence via a factor $\tau_{\rm d}/\tau_{\rm q} \propto
\sigma_{\perp}d^2$ -- see e.g., equations (A6)--(A11) in
\cite{Krekhov07}. Nevertheless, in the conductive regime, $U_{\rm
c}$ is thickness independent in the lowest order time Fourier
approximation for the experimentally most relevant ($10 -100 \mu$m)
thickness range, where $\tau_{\rm d} \gg \tau_{\rm q}$
\cite{Bodenschatz88}. In contrast, in the dielectric regime an
analogous approximation gives $U_{\rm c} \propto d$.

At typical ($d \geq 10\mu$m) cell thicknesses the conductive regime
appears at low frequencies and the dielectric rolls are observable
at higher $f$. The change of pattern type occurs at $f_{\rm c}$
which depends significantly on $\sigma_{\perp}$ as well as on $d$.
Therefore, in order to extend the frequency range of the dielectric
regime, thus to shift $f_{\rm c}$ to lower values, one has to reduce
either the cell thickness or the electric conductivity or both. When
both $\sigma_{\perp}$ and $d$ are small enough, the conductive
regime does not occur at all.

The SM regards NLCs as regular dielectrics, where the source of the
electric polarization is the electric field. In NLCs, however,
flexoelectric polarization \cite{Meyer69,deGennes,Chandrasekhar} may
appear even in the absence of the electric field, if the director
field is distorted. Flexoelectricity produces an additional
contribution to the charge distribution and to the electric torque
acting on the director. It is also known for a long time that under
large enough dc voltage, flexoelectricity may distort the director
field in the planar geometry resulting in a non-convective striped
pattern oriented parallel with the initial {\bf n} \cite{Vistin74}.
The threshold voltage and the wavenumber of this non-convective
pattern, as well as the frequency dependence of the threshold
characteristics have been calculated \cite{Bobylev77,Barnik77}.

The effect of flexoelectricity on EC has also been studied and it
has been shown that for the commonly used nematic material
parameters, and for a typical sample thickness of $d \gtrsim 10
\mu$m, in the usually studied (not too low) frequency range of
conductive s-EC, flexoelectricity has no significant influence on
the characteristics at the onset of instability
\cite{Kramer89,thom,Madhu}. That gives the reason why the
contribution of the flexoelectric polarization has usually been
neglected in the SM. On the other hand, it has also been shown
that when applying a dc voltage, the EC threshold becomes
independent of $d$ and $\sigma_{\perp}$ \cite{Kramer89}. Moreover,
flexoelectricity leads to an appreciable reduction of the dc EC
threshold and it also influences the direction of the rolls
\cite{thom}. Obviously, there must be an ac frequency range in
which the effect of flexoelectricity diminishes with the increase
of the frequency. This frequency range, however, has not yet been
studied in details.

In the extended SM (i.e., with the inclusion of the flexoelectric
polarization) two additional parameters, the flexoelectric
coefficients $e_1$ and $e_3$ \cite{Meyer69,Zimmer91} come into play.
An analysis of the nematohydrodynamic equations linearized around
the basic state has proven that flexoelectricity establishes a
coupling between the conductive and dielectric modes introduced
above \cite{Zimmer91,Krekhov07} resulting in a complex time
dependence. The solutions still can be classified according to their
spatiotemporal parity \cite{Krekhov07}. The even parity solution has
mode I coupled to mode IV (mode I+IV), and corresponds to the
"conductive regime", while in the odd parity "dielectric" solution
mode III couples to mode II (mode II+III) \cite{remark}.

What has been said above, applies to substances with material
parameters satisfying $\varepsilon_a < 0$, $\sigma_a > 0$ or
$\varepsilon_a >0$,  $\sigma_a < 0$ with the appropriate boundary
conditions \cite {Buka05}. In contrary, in case of nematics with
$\varepsilon_a < 0$ and $\sigma_a < 0$ \cite{Buka05,deGennes}, where
the feedback loop remains negative for all voltages, the
Carr-Helfrich mechanism excludes the existence of EC patterns.
Nonetheless, a convection roll pattern has long ago been observed in
such compounds in ac electric field \cite{Goscianski75,Blinov79} and
has been reconsidered systematically recently in a few nematics
\cite{Ela,Toth07,Kumar07}. Some basic characteristics of these
patterns, like the orientation of the rolls (nearly parallel with
the initial {\bf n}), the contrast, the frequency dependence of
$U_{\rm c}$ and that of the critical wave number have been found
considerably different from those of the s-EC. Since this kind of
pattern formation is not predicted by the classical SM (without the
incorporation of the flexoelectricity), it has been called
nonstandard electroconvection (ns-EC) \cite{Ela}. Recently,
convection patterns with similar properties have been detected in
nematics built up from bent-core molecules
\cite{Wiant05,Tamba07,Stannarius07} that may exhibit giant
flexoelectricity \cite{Harden06}.

These experiments have triggered a recent reconsideration of the
role of flexoelectricity within the extended SM for a wider range of
material parameters. It has been shown that in some cases
flexoelectricity cannot be disregarded. At high frequencies, in the
dielectric regime a nonzero $e_1$ and $e_3$ leads to a considerable
decrease of $U_{\rm c}$ \cite{Krekhov}. What is even more important,
the flexoelectric contribution to the charge separation yields
finite instability threshold for planar nematics with $\sigma_a <0$
and $\varepsilon_a <0$, thus giving an explanation for ns-EC
\cite{Krekhov07}.

Special behavior can be expected (some might go beyond the
applicability limit of the SM) in the parameter ranges where the
period of the driving frequency $1/f$ becomes comparable with one of
the characteristic times of the system $\tau_{\rm d}$, $\tau_{\rm
q}$ or $\tau_{\rm v}$. The majority of the experimental and
theoretical studies of EC so far apply to relatively thick cells of
medium conductivity, thus they do not extend to this range. We aim
to fill this gap and we also show that flexoelectricity plays a
crucial role in those ranges.

In the present paper we report on experimental studies and numerical
simulations of EC aiming to approach the characteristic times of the
system by $1/f$, especially concentrating on $\tau_{\rm d}$ and in
some special cases on $\tau_{\rm q}$ ($\tau_{\rm v}$ is usually very
short, to approach it would require very thick cells and high
driving frequencies, which would result in high threshold voltages
not accessible experimentally). We focus on the frequency dependence
of the threshold voltage $U_{\rm c}(f)$. The measurements have been
carried out on five different nematic compounds: four of them
exhibit s-EC, while in the fifth one, both s-EC and ns-EC patterns
could be observed. The numerical analysis has been based on the
extended standard model of EC which includes the flexoelectric
effects \cite{Zimmer91,Krekhov07}.

The structure of the paper is organized according to the pattern
type in the low frequency range. After introducing the substances,
the setup, and some details of the numerical analysis in Section
\ref{sec:setup}, Section \ref{sec:condsEC} is devoted to results on
compounds in which conductive s-EC has been detected at low
frequencies. In Section \ref{sec:dielsEC} we discuss samples which
exhibit dielectric s-EC at low $f$, while Section \ref{sec:nsEC}
describes the ns-EC instability. Finally, we conclude the paper with
the discussion in Section \ref{sec:concl}.

\section{\label{sec:setup}Substances, experimental set-up, and details of calculations}

Five different NLCs have been used in the measurements. The three
commercial nematic mixtures, Phase 5, Phase 5A and Phase 4 (from
Merck \& Co., Inc.) as well as the
4-methoxy-benzylidene-4'-n-butyl-aniline (MBBA) have $\sigma_a>0$
and $\varepsilon_a<0$; so they exhibit s-EC upon application of an
electric voltage. All measurements presented here on these compounds
have been performed at $T=30^\circ$C.

The fifth studied compound was the
4-n-octyloxy-phenyl-4-n'-heptyloxy-benzoate (\textbf{8/7})
\cite{Kresse80} which has the phase sequence: isotropic -- 92
$^{\circ}$C -- nematic -- 72.5 $^{\circ}$C -- smectic C --
62$^{\circ}$C -- crystalline. It has $\varepsilon_a < 0$ over the
whole nematic temperature range, but shows a sign inversion from
$\sigma_a <0$ to $\sigma_a >0$ as the temperature is increased
toward the clearing point. Consequently, in this particular compound
both ns-EC and s-EC could be investigated (although at different
temperatures).

Our aim is to study the EC onset behavior in the vicinity of the
characteristic relaxation times of the system;  in order to approach
that we have varied the cell thickness and the conductivity to tune
$\tau_{\rm d}$ and $\tau_{\rm q}$.

The nematic liquid crystals have been enclosed between two parallel
glass plates coated with etched transparent indium tin oxide (ITO)
electrodes. Rubbed polyimide has been used to obtain a planar
alignment. Cells in the thickness range from $d \approx 3 \mu$m to
$d = 40 \mu$m have been prepared. A spectrophotometer has been used
for measuring $d$. It also enabled a systematic mapping of the
thickness of the empty cell, thus determining its total variation
throughout the whole active area (region between electrodes) of the
sample. Typically, we have found a variation of $\pm 0.5 \mu$m. The
conductivity range extended from $10^{-9} ( \Omega {\rm m})^{-1}$ to
$10^{-7}( \Omega {\rm m})^{-1} $. The direction of the director at
the surfaces is chosen as the $x$-axis. An electric field across the
sample (along the $z$-axis) has been generated by applying a
sinusoidal ac electric voltage of frequency $f$ and amplitude
$\sqrt{2} U$ to the electrodes. The cells have been placed into an
oven (an Instec hot-stage) thermostatted within $\pm 0.05
^{\circ}$C.

EC patterns have been studied with polarizing microscopy using
either the shadowgraph (single polarizer) technique or two crossed
(or nearly crossed) polarizers. The images have been recorded with
a CCD camera, digitized by a frame grabber with a resolution of at
least $768 \times 576$ pixels and 24 bit color depth and saved for
further processing/analysis.

Numerical simulations have aimed to compute the frequency dependent
threshold voltage $U_{\rm c}(f)$ for the different EC patterns. The
extended SM (with flexoelectricity included) has been used. Being
interested in the onset behavior, linear stability analysis has been
carried out as described in detail in \cite{Krekhov07}. Strong
anchoring of the director and no slip condition for the flow have
been assumed at the bounding plates, which have been ensured by a
Galerkin method. The field variables have been expanded into sets of
functions that vanish at the boundaries; the time periodicity has
been guaranteed by Fourier expansion. Expansions in space and time
have been truncated at the fourth and the seventh modes,
respectively, which has been proved sufficient to obtain accuracy of
better than $1\%$ in $U_{\rm c}(f)$. This has been occasionally
checked by increasing systematically the number of modes in space
and time up to ten and fifteen, respectively, and monitoring the
changes in $U_{\rm c}$. The growth rates $\mu(\mathbf{q},U)$ for
spatial fluctuations have been calculated as a function of the
applied voltage. The minimum of the neutral surface
[$\mu(\mathbf{q},U)=0$] provided the threshold voltage $U_{\rm c}$.
The numerical calculations have been carried out with a Fortran code
developed at the University of Bayreuth, using the same set of known
material parameters (permittivities, conductivities, elastic moduli,
viscosities), as for earlier calculations with the SM (for Phase 5
and Phase 5A in \cite{Treiber97}, and for MBBA in
\cite{Bodenschatz88,Hertrich92,Zhou06}). The original measurements
of these parameters have been reported in
\cite{Treiber97,Merck,Graf92} for Phase 5 and in
\cite{deJeu76,Rondelez71,Sprokel73,Sinclair76,Kneppe82,Gahwiller71}
for MBBA.

The splay and bend flexoelectric coefficients, $e_1$ and $e_3$,
should also be known, however, they are not easily measurable.
Though various measuring techniques have been developed during the
past decades, there is still a big controversy about the value and
the sign of $e_1$ and $e_3$. The data obtained by different methods
vary considerably even for the most studied nematic material, MBBA:
$e_1+e_3$ ranges from $-2.3$pC/m \cite{Blinov88} to $-(54 \pm
10)$pC/m \cite{Takahashi98}, and $e_1-e_3$ from $(3.3 \pm 0.7)$pC/m
\cite{Dozov82} to $(14 \pm 1)$pC/m \cite{Takahashi98}. Up to now,
only the sign of the sum of flexoelectric coefficients seems to be
above dispute for MBBA: it appears to be negative ($e_1+e_3 < 0$).
For the other four compounds used in our experiments no data on the
flexoelectric coefficients are at all currently available. It seems
though, based on measurements on a number of (other) nematics, that
at least the order of magnitude of $\sim 10$pC/m is established as a
customary value for both $e_1$ and $e_3$ \cite{Petrov01}. As a
consequence, during the numerical simulations we have regarded $e_1$
and $e_3$ as adjustable parameters, to be chosen to provide the best
fit with the experiments.

\section{\label{sec:condsEC}Conductive s-EC at low frequencies and dielectric s-EC at high frequencies}

We start the discussion of the experimental results with systems
exhibiting s-EC, where by choosing $d$ and $\sigma_{\perp}$
appropriately, the frequency range of the ac field covers
$1/\tau_{\rm d}$ and in most cases $1/\tau_{\rm q}$ as well.

The first case we show is electroconvection in Phase 5 with  $d=(3.4
\pm 0.5)\mu$m and $\sigma_{\perp}=8.2 \times 10^{-8} (\Omega {\rm
m})^{-1}$. In this case $1/\tau_{\rm d} \approx 60$Hz, $1/\tau_{\rm
q} \approx 1800$Hz, and $1/\tau_{\rm v} \approx 2 \times 10^6$Hz.

\begin{figure}[h!t]
\includegraphics[width=8.5cm]{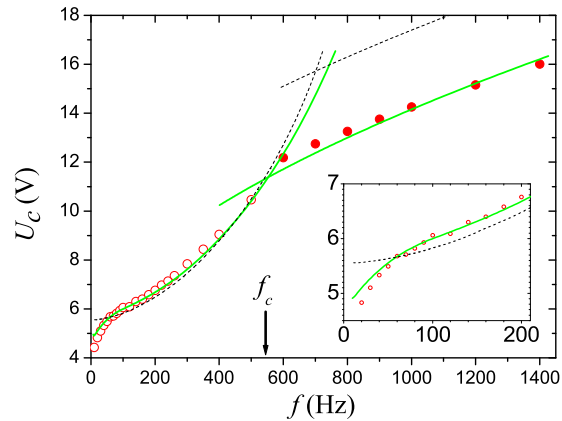}\\
\caption{(Color online) Frequency dependence of the s-EC threshold
voltage $U_{\rm c}$ measured in Phase 5 with $d=(3.4 \pm 0.5)\mu$m
-- conductive s-EC (circles) and dielectric s-EC (bullets). The
solid and the dashed lines are the results of the numerical analysis
with and without the flexoelectric effect, respectively. The inset
is the magnification of the low frequency range. The arrow indicates
the crossover frequency $f_{\rm c}$.} \label{Ph5d3_9mall}
\end{figure}

Threshold measurements are presented in Fig. \ref{Ph5d3_9mall}.
Below $f_{\rm c} \approx 550$Hz conductive s-EC (open circles) has
been detected at onset in the form of traveling oblique rolls and
traveling normal rolls. Above $f_{\rm c}$ traveling dielectric s-EC
(bullets) has been observed.

A novel feature of the $U_{\rm c}(f)$ curve has been detected in the
low frequency range, $20$Hz$\lesssim f \lesssim 60$Hz, where the
threshold strongly decreases by lowering the frequency, in contrast
to thick ($d \geq 10 \mu$m) cells where $U_{\rm c}$ remains almost
constant in the same frequency range.

Numerical calculations have been performed with the Phase 5
parameter set (see Table~\ref{table1}), using the flexoelectric
coefficients $e_1$ and $e_3$ as fitting parameters. In
Fig.~\ref{Ph5d3_9mall} the results are plotted both without the
flexoelectric effect ($e_1=e_3=0$, dashed lines, representing mode
I and mode III) and with flexoelectric coefficients of
$e_1=-26.5$pC/m and $e_3=-23.6$pC/m (solid lines, mode I+IV and
mode II+III). For the conductive s-EC, the calculation without the
flexoelectric effect does not show the bending down at low
frequencies while with the $e_1$ and $e_3$ values given above a
good match with the experimental data has been obtained (see the
two curves in the inset of Fig.~\ref{Ph5d3_9mall}).

\begin{table*}[h!t]
\caption{\label{table1} Material parameters of Phase 5 and MBBA at
$T=30^{\circ}C$. Phase 5 parameters have also been used in numerical
calculations for Phase 5A.}
\begin{tabular}{|c|c|c|c|}
\hline
     Parameter & Unit & Phase 5 &  MBBA \\
     \hline
          \hline
     $K_{11}$ & $10^{-12}$N & 9.8 \cite{Treiber97} & 5.89 \cite{deJeu76}\\
     $K_{22}$ & $10^{-12}$N & 4.6 \cite{Treiber97} & 3.7 \cite{deJeu76}\\
     $K_{33}$ & $10^{-12}$N & 12.7 \cite{Treiber97} & 7.56 \cite{deJeu76}\\
     \hline
  $\sigma_a/\sigma_{\perp}$ & -- & 0.69 \cite{Treiber97} & 0.5 \cite{Rondelez71}\\
   \hline
   $\varepsilon_{\perp}$ & -- & 5.25 \cite{Treiber97} & 5.18 \cite{Rondelez71}\\
   $\varepsilon_a$ & -- & $-0.184$ \cite{Treiber97,Merck} & -0.48 \cite{Rondelez71}\\
\hline
   $\alpha_1$ & $10^{-3}$Ns/m$^2$ & $-39$ \cite{Treiber97} & -14.1 \cite{Kneppe82}\\
   $\alpha_2$ & $10^{-3}$Ns/m$^2$ & $-109.3$ \cite{Graf92} & -80.0 \cite{Kneppe82}\\
   $\alpha_3$ & $10^{-3}$Ns/m$^2$ & 1.5 \cite{Graf92} & -1.513 \cite{Kneppe82}\\
   $\alpha_4$ & $10^{-3}$Ns/m$^2$ & 56.3 \cite{Graf92} & 64.4 \cite{Kneppe82}\\
   $\alpha_5$ & $10^{-3}$Ns/m$^2$ & 82.9 \cite{Graf92} & 57.2 \cite{Kneppe82}\\
   $\alpha_6$ & $10^{-3}$Ns/m$^2$ & -24.9 \cite{Graf92} & -24.4 \cite{Kneppe82}\\
\hline
   $e_1$ & pC/m & $-26.5$ (fit) & $-14.5$ (fit)\\
   $e_3$ & pC/m & $-23.6$ (fit) & $-20.5$ (fit)\\
\hline
\end{tabular}
\end{table*}

\begin{figure}[h!t]
\includegraphics[width=8.5cm]{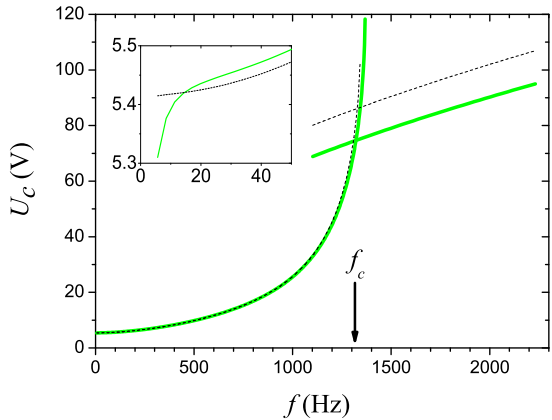}\\
\caption{(Color online) Numerical calculations for the frequency
dependence of the s-EC threshold voltage $U_{\rm c}$ in a $d=20\mu$m
sample of Phase 5 with $\sigma_{\perp}=8.2 \times 10^{-8} (\Omega
{\rm m})^{-1}$. Dashed lines are the conductive s-EC and the
dielectric s-EC solutions with $e_1=e_3=0$, while the solid lines
are the solutions with $e_1=-26.5$pC/m and $e_3=-23.6$pC/m. The
inset is the magnification of the low frequency range. The arrow
indicates the crossover frequency $f_{\rm c}$.}
\label{UvsfPh5micron20}
\end{figure}

Calculations have been performed for a $d=20\mu$m cell as well,
using the same parameters from Table \ref{table1} (providing
$1/\tau_{\rm d} \approx 2$Hz, $1/\tau_{\rm q} \approx 1800$Hz, and
$1/\tau_{\rm v} \approx 70 \times 10^3$Hz). The results are plotted
in Fig.~\ref{UvsfPh5micron20}. One can immediately see that the
resulting $U_{\rm c}(f)$ for $e_1=e_3=0$ (mode I) nearly coincides
with the one incorporating the flexoelectric effect  (mode I+IV)
almost in the whole conductive frequency range, in accordance with
earlier conclusions \cite{Kramer89}. Nevertheless a closer look at
the curves (see the inset of Fig.~\ref{UvsfPh5micron20} with a blow
up of the plot) shows that switching on the flexoelectricity results
in the bending down of the threshold here too, similarly to the case
of low $d$, only the effect occurs here at much lower frequencies.

The bending down of $U_{\rm c}$ at low $f$ when flexoelectricity is
included can be understood from the earlier theoretical results
\cite {thom} where it has been shown that flexoelectricity decreases
the dc threshold voltage considerably (for MBBA the reduction is
about 25\%). It could also be proved that in this dc limit none of
the dimensionless equations of SM depend on $\sigma_{\perp} d^2$,
hence the dc threshold is independent of $d $ and $\sigma_{\perp}$
\cite{Kramer89}. In case of ac driving there is a wide frequency
range (in the middle of the conductive regime) where
flexoelectricity has much less influence on $U_{\rm c}$; it leaves
the ac threshold almost unaltered, at least for not too thin cells.
This means that at increasing $f$ the contribution from the
flexoelectric charge separation (important at dc) fades away
compared to that of the Coulomb charge density in a narrow frequency
range at low $f$. The character of this transition from dc to ac
driving has, however, not been studied rigorously either
experimentally or theoretically so far.

\begin{figure}[h!t]
\includegraphics[width=8.5cm]{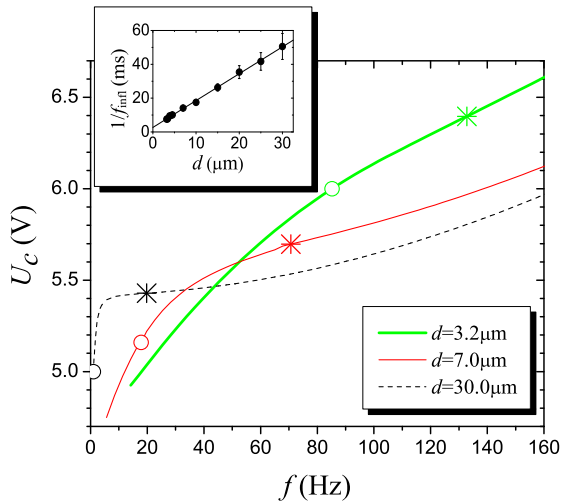}\\
\caption{(Color online) Numerically calculated frequency dependence
of the s-EC threshold voltage $U_{\rm c}$ for samples of various
thicknesses in Phase 5 with $\sigma_{\perp}=8.2 \times 10^{-8}
(\Omega {\rm m})^{-1}$, $e_1=-26.5$pC/m and $e_3=-23.6$pC/m. The
circles and the star symbols on the curves denote $1/\tau_{\rm d}$
and the inflection point at $f_{\rm infl}$, respectively. Inset: the
thickness dependence of $1/f_{\rm infl}$. } \label{UvsfPh5thickness}
\end{figure}

In order to understand better the behavior of $U_{\rm c}$ we have
carried out further simulations, carefully analyzing the low
frequency range for a series of thicknesses when flexoelectricity is
included. Results are shown in Fig.~\ref{UvsfPh5thickness} for three
thicknesses. The figure indicates that the transition from ac to dc
(the bending down of $U_{\rm c}$ to its dc value) occurs for any
thickness as expected; only the effect is compressed into a very
narrow frequency range for large thicknesses. The corresponding
values of $1/\tau_{\rm d}$ are also indicated in
Fig.~\ref{UvsfPh5thickness} (circles). Their values fall into those
frequency ranges (different for each $d$) in which the strong
bending down of $U_{\rm c}$ occurs. We note, that according to its
definition, $\tau_{\rm d}$ is the relaxation time of a homogeneously
deformed director state (wavenumber $|{\mathbf q}|=0$). The
relaxation time of the pattern, $1/\tau_{\rm p}$, is shorter,
because the growth/decay rate ($\mu =1/\tau_{\rm p}$) of a periodic
($|{\mathbf q}| \neq 0$) distortion increases with $|{\mathbf q}|$
\cite{Eber04,Pesch06}. Therefore $1/\tau_{\rm p}>1/\tau_{\rm d}$
would be a more appropriate characteristic frequency of the system
than $1/\tau_{\rm d}$. A rigorous quantitative comparison of
$1/\tau_{\rm p}$ with $f$ is, however, problematic since $|{\mathbf
q}|$ and hence $\tau_{\rm p}$ are frequency dependent.

The crossover between the regimes where the Coulomb charge
separation or the flexoelectric charge separation dominates, is
reflected in the curvature of the $U_{\rm c}(f)$ curves. It is
convenient to use the location of the inflection point at $f_{\rm
infl}$ (where the curvature changes sign, indicated by stars in
Fig.~\ref{UvsfPh5thickness}) as a quantitative measure. As one sees
values of $f_{\rm infl}$ are close to $1/\tau_{\rm d}$, but
interestingly, $1/f_{\rm infl}$ is linear with $d$ (see in the inset
of Fig.~\ref{UvsfPh5thickness}), in contrast to the $d^2$ dependence
of $\tau_{\rm d}$.

Finally, we mention that for the high frequency end of the
conductive regime the influence of flexoelectricity becomes again
noticeable though less dramatic. It considerably enhances the
$d$-dependence (see in Fig.~\ref{UvsfPh5thickness}) which would be
almost negligible at $e_1=e_3=0$, especially for higher thicknesses.
Furthermore it shifts $f_{\rm c}$ to lower frequencies as seen in
Fig.~\ref{Ph5d3_9mall}.

The shift of $f_{\rm c}$ is mainly due to the fact that in the
dielectric regime of s-EC the flexoelectric effect decreases the
threshold voltage drastically -- compare the dashed curve
($e_1=e_3=0$, mode III) and the solid line ($e_1=-26.5$pC/m and
$e_3=-23.6$pC/m, mode II+III) for the dielectric solutions above
$f_{\rm c}$ in Figs.~\ref{Ph5d3_9mall} and \ref{UvsfPh5micron20}.
The reduction reaches about $16\%$ even in the relatively thick
($d=20\mu$m) sample, which agrees with recent findings for MBBA
\cite{Krekhov}. Consequently, flexoelectric effects cannot be
neglected in case of dielectric s-EC even in the thick samples. We
also mention here that no peculiar effect on $U_{\rm c}(f)$ is
observed in the vicinity of $1/\tau_{\rm q} \approx 1800$Hz (see
Fig. \ref{UvsfPh5micron20}).

Summarizing the results above one can conclude that with the above
mentioned values of $\sigma_{\perp}$, $e_1$ and $e_3$, an excellent
agreement has been found between the experimental results (symbols)
and the numerical calculations (solid lines in
Fig.~\ref{Ph5d3_9mall}) for Phase 5 in the whole experimentally
covered frequency range. Note that the fitted $e_1$ and $e_3$ values
are realistic in a sense that they are of the same order of
magnitude as those measured for other nematics.

\begin{figure}[h!t]
\includegraphics[width=8.5cm]{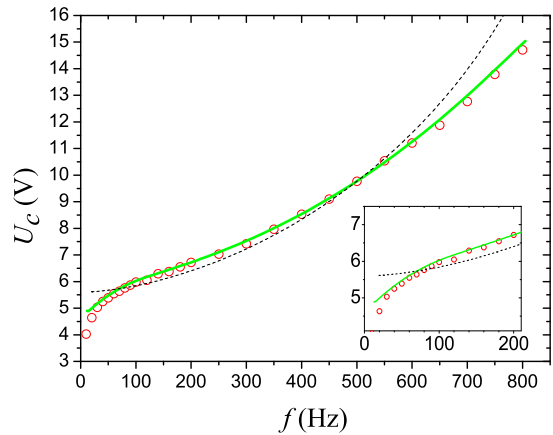}\\
\caption{(Color online) Frequency dependence of the threshold
voltage $U_{\rm c}$ in the conductive regime for a sample of Phase
5A with $d=3.1\mu$m. Circles stand for experimental data. The solid
line represents the $U_{\rm c}(f)$ curve calculated numerically with
Phase 5 parameters (same as in Figure 1, with nonzero $e_1$ and
$e_3$) and with $\sigma_{\perp}=9.2 \times 10^{-8} (\Omega {\rm
m})^{-1}$. The dashed line is the calculated solution for
$e_1=e_3=0$. The inset is the magnification of the low frequency
range.} \label{Ph5Acomp}
\end{figure}

Similar results have also been obtained on Phase 5A and MBBA samples
in which only the conductive regime of s-EC has been detected. Phase
5A has been investigated in a sample of $d=3.1\mu$m. At dc voltage
as well as at very low frequencies (up to $f \approx 3$Hz) static
stripe patterns, the so called flexoelectric domains
\cite{Bobylev77,Marinov01} have been observed at a threshold $U_{\rm
f}$. Above $f=3$Hz only conductive s-EC patterns have been seen: up
to $f=40$Hz stationary oblique rolls and above $f=40$Hz traveling
oblique rolls and traveling normal rolls. The measured frequency
dependence of the threshold voltage is presented in Fig.
\ref{Ph5Acomp} by open circles. Below $f \approx 50$Hz it appears
that $U_{\rm c}$ decreases considerably as approaching $f=0$ (see
the inset in Fig.~\ref{Ph5Acomp}), in a similar fashion as seen for
the Phase 5 in Fig.~\ref{Ph5d3_9mall}. Again the bending down occurs
in the frequency range of $1/\tau_{\rm d}$.

Phase 5A has the same chemical composition as Phase 5 except that
Phase 5A is doped with some ionic salt in order to increase the
electric conductivity. Therefore, in the numerical calculations
for Phase 5A we have used the material parameters of Phase 5
(given in Table~\ref{table1}) with $\sigma_{\perp}=9.2 \times
10^{-8} (\Omega {\rm m})^{-1}$, and the flexoelectric coefficients
$e_1=-26.5$pC/m and $e_3=-23.6$pC/m that have been obtained from
the best fit in Fig.~\ref{Ph5d3_9mall}. The calculated threshold
curve for conductive s-EC (solid line in Fig.~\ref{Ph5Acomp})
agrees very well with the experimental data in the whole frequency
range.

For reference, in Fig.~\ref{Ph5Acomp} we have also plotted the
conductive s-EC threshold $U_{\rm c}(f)$ without the flexoelectric
effect (i.e., for $e_1=e_3=0$; dashed line). The same conclusions
can be drawn here as those for Phase 5: the $U_{\rm c}(f)$ curve
with $e_1=e_3=0$ deviates significantly and systematically from
that with nonzero $e_1$ and $e_3$ (and also from the experimental
data), especially in the low frequency range (below $f \approx
50$Hz) and at high frequencies (above 600Hz).

\begin{figure}[h!t]
\includegraphics[width=8.5cm]{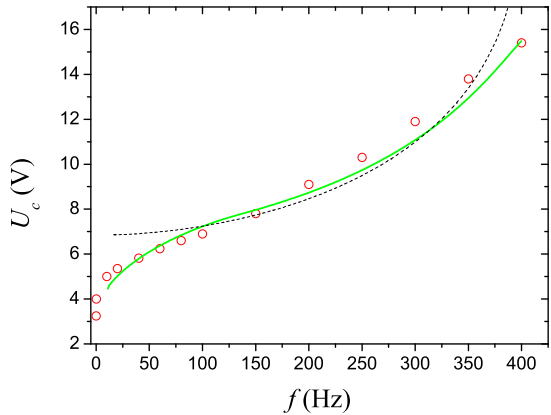}\\
\caption{Frequency dependence of the conductive s-EC threshold
voltage $U_{\rm c}$ measured on a sample of MBBA with $d=3.2\mu$m
(circles). The solid line represents the $U_{\rm c}(f)$ curve
calculated numerically with MBBA parameters from Table \ref{table1}
and with $\sigma_{\perp}=9 \times 10^{-8} (\Omega {\rm m})^{-1}$,
$e_1=-14.5$pC/m and $e_3=-20.5$pC/m. The dashed line is the solution
for the same parameter set, but with $e_1=e_3=0$.} \label{MBBAcomp}
\end{figure}

Up to now the most studied nematic EC material is presumably MBBA
which we have investigated too, in a cell of $d=3.2\mu$m. At dc
voltage neither flexoelectric domains, nor periodic EC patterns have
been detected up to $U=80$V. However, at and above $f=0.1$Hz a
conductive s-EC pattern has emerged. Frequency dependence of the
threshold voltage measured in this thin layer of MBBA is shown in
Fig.~\ref{MBBAcomp} (circles). Again, below $f \approx 30$Hz $U_{\rm
c}(f)$ bends down considerably while approaching $f \rightarrow 0$,
in a similar fashion as seen for Phase 5 and Phase 5A.

The MBBA parameter set from Table \ref{table1} has been used in the
numerical calculations. The solid line in Fig. \ref{MBBAcomp}
obtained by the simulations for $e_1=-14.5$pC/m, $e_3=-20.5$pC/m and
$\sigma_{\perp}=9 \times 10^{-8} (\Omega {\rm m})^{-1}$ represents a
fairly good match with the experimental data, in contrast to the
curve with $e_1=e_3=0$ (dashed line in Fig. \ref{MBBAcomp}) which
deviates significantly. One has to note here that the fitted values
of $e_1$ and $e_3$ fall in the mid-range of measured values for MBBA
\cite{Petrov01}, and in particular agree reasonably well with the
values $|e_1|=15$pC/m and $|e_3|=30$pC/m reported in
\cite{Scaldin85,Scaldin90}.

\section{\label{sec:dielsEC}Dielectric s-EC at low frequencies}

With a careful preparation procedure we were able to produce thin
($d=3.4\mu$m) Phase 5 samples with a conductivity as low as about
$\sigma_{\perp} \approx 1 \times 10^{-8} (\Omega {\rm m})^{-1}$.
These samples exhibited convection-free flexoelectric domains at a
threshold $U_{\rm f}$ at dc and at very low frequency (up to
$f=7.5$Hz) ac driving, similarly to the situation in Phase 5A. Above
$f=7.5$Hz, in a narrow frequency range (up to $f\approx 20$Hz)
flexoelectric domains and traveling dielectric s-EC oblique rolls
coexist with slightly different thresholds. At higher frequencies
dielectric s-EC could be detected in the form of traveling oblique,
or traveling normal rolls.

\begin{figure}[h!t]
\includegraphics[width=8.5cm]{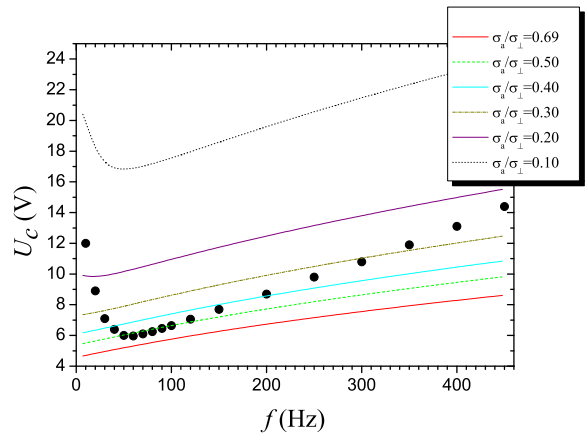}\\
\caption{(Color online) Frequency dependence of the dielectric
threshold voltage $U_{\rm c}$ measured in a $d=3.4\mu$m sample of
Phase 5 (bullets). Lines represent results of numerical calculations
for Phase 5 with $e_1=-26.5$pC/m, $e_3=-23.6$pC/m, and
$\sigma_{\perp}=1 \times 10^{-8} (\Omega {\rm m})^{-1}$ for a series
of $\sigma_a/\sigma_{\perp}$ values.} \label{Uvfdielvarycondaniz}
\end{figure}

The frequency dependence of $U_{\rm c}$ in this sample (depicted
with bullets in Fig.~\ref{Uvfdielvarycondaniz}) is, however, rather
unusual and surprising, considering that both the SM and previous
experiments on dielectric s-EC provided a square root-like $U_{\rm
c}(f)$ function. Instead, here $U_{\rm c}(f)$ bends up as $f
\rightarrow 0$, i.e. it is a non-monotonic function with an
expressed minimum at $f_{\rm min}$ in the frequency range of
$1/\tau_{\rm d}$. Moreover, at higher frequencies (above the
minimum) $U_{\rm c}(f)$ is linear within the experimental error.
Performing numerical simulations with the Phase 5 parameter set in
Table~\ref{table1} and with $\sigma_{\perp} = 1 \times 10^{-8}
(\Omega {\rm m})^{-1}$, we have found that the calculated dielectric
threshold curve $U_{\rm c}(f)$ (shown as the solid line with the
lowest $U_{\rm c}$ in Fig.~\ref{Uvfdielvarycondaniz}) is monotonic,
has a smaller slope (above $60$Hz) and provides lower thresholds
(especially at high $f$) than found experimentally.

In order to test whether the extended SM is {\it able} at all to
provide an $U_{\rm c}(f)$ dependence with at least a {\it
qualitative} similarity to the measured data, we have performed
simulation experiments. First we have checked the influence of the
conductivity systematically in a wide range of $10^{-12} (\Omega
{\rm m})^{-1} \leq \sigma_{\perp} \leq 10^{-8} (\Omega {\rm
m})^{-1}$, where only dielectric s-EC is expected. We could verify
that the calculated $U_{\rm c}(f)$ remains monotonic in this
$\sigma_{\perp}$ range, and ${\rm d}U_{\rm c}/{\rm d}f$ is not
affected significantly. Consequently, the novel frequency dependence
of the threshold (bending up) cannot be interpreted by a pure
reduction of the conductivity.

We have also varied the anisotropy of the conductivity motivated by
recent numerical results \cite{Krekhov07} indicating that a decrease
of $\sigma_a/\sigma_{\perp}$ increases the slope ${\rm d}U_{\rm
c}/{\rm d}f$ and tends to make $U_{\rm c}(f)$ linear. In
Fig.~\ref{Uvfdielvarycondaniz} we present the calculated threshold
curves for five additional $\sigma_a/\sigma_{\perp}$ values, leaving
all other parameters in Table~\ref{table1} unaltered. It is seen
that $U_{\rm c}(f)$ as well as ${\rm d}U_{\rm c}/{\rm d}f$ shift to
higher values with the decrease of $\sigma_a/\sigma_{\perp}$,
moreover, $U_{\rm c}(f)$ becomes non-monotonic for
$\sigma_a/\sigma_{\perp} \leq 0.2$ with a bending up of $U_{\rm
c}(f)$ at low frequencies. Note that the curve calculated for
$\sigma_a/\sigma_{\perp}=0.1$ reproduces qualitatively all features
of the experimental $U_{\rm c}(f)$, except that it is shifted to
higher voltages.

\begin{figure}[h!t]
\includegraphics[width=8.5cm]{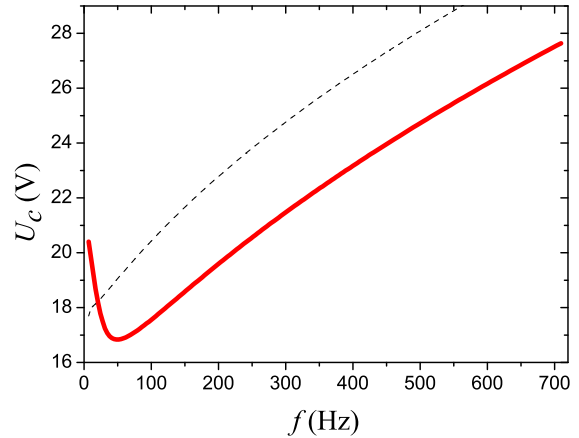}\\
\caption{(Color online) Numerical calculations of the threshold
$U_{\rm c}(f)$ for Phase 5 with $\sigma_a/\sigma_{\perp}=0.1$,
$\sigma_{\perp}=1 \times 10^{-8} (\Omega {\rm m})^{-1}$, and
$d=3.4\mu$m: dashed line is dielectric s-EC with $e_1=e_3=0$; solid
line is the dielectric mode with $e_1=-26.5$pC/m, $e_3=-23.6$pC/m.}
\label{Ph5cleansim}
\end{figure}

To demonstrate the role of flexoelectricity in the novel frequency
dependence of the dielectric regime, in Fig.~\ref{Ph5cleansim} we
have plotted $U_{\rm c}(f)$ calculated with
$\sigma_a/\sigma_{\perp}=0.1$ both for mode III (i.e., without
flexoelectricity, dashed line) and for mode II+III (i.e., with
flexoelectricity, solid line). Obviously, when flexoelectric effects
are neglected, the threshold is even higher, $U_{\rm c}(f)$
decreases monotonously when $f\rightarrow 0$ and the bending up is
not reproduced, so there is no agreement with the experimental data
even on a qualitative level. On the other hand, if one takes into
account the flexoelectric coefficients given in Table~\ref{table1},
an acceptable qualitative agreement is achieved: the curve has a
minimum at about the same $f_{\rm min}$ frequency where the
experimental data, and for $f > f_{\rm min}$ the steepness ${\rm
d}U_{\rm c}/{\rm d}f$ is roughly the same as that in the
experiments. Therefore, the bending up can undoubtedly be attributed
to flexoelectricity, and $f_{\rm min}$ is a natural indicator for
the frequency range where the crossover between dominantly
flexoelectric and dominantly Coulomb charge separation mechanisms
occur.

Finally we mention that (similarly to the case discussed for Fig.
\ref{UvsfPh5micron20}) no peculiar effect of flexoelectricity on
$U_{\rm c}(f)$ is observed in the vicinity of $1/\tau_{\rm q}
\approx 220$Hz (see Figs. \ref{Uvfdielvarycondaniz} and
\ref{Ph5cleansim}).

\begin{figure}[h!t]
\includegraphics[width=8.5cm]{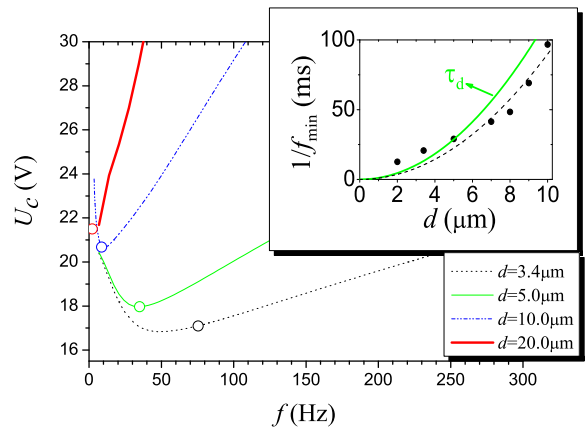}\\
\caption{(Color online) Frequency dependence of the dielectric s-EC
threshold, calculated with Phase 5 parameters for
$\sigma_a/\sigma_{\perp}=0.1$ and with $e_1=-26.5$pC/m,
$e_3=-23.6$pC/m, $\sigma_{\perp}=1 \times 10^{-8} (\Omega {\rm
m})^{-1}$ for different sample thicknesses $d$. Circles denote
$1/\tau_{\rm d}$ for each $d$. In the inset the thickness dependence
of $1/f_{\rm min}$ (bullets) and that of $\tau_{\rm d}$ (solid line)
is presented together with the best fit $1/f_{\rm min} \propto d^2$
(dashed line).} \label{Uvfdielddependence}
\end{figure}

Systematic numerical calculations at $\sigma_a/\sigma_{\perp}=0.1$
have shown that while keeping $\sigma_{\perp}$ (and thus $\tau_{\rm
q}$) constant, $f_{\rm min}$ shifts to higher $f$ with the decrease
of $d$ (i.e., of $\tau_{\rm d}$) -- see Fig.
\ref{Uvfdielddependence}. Despite of the fact that $1/f_{\rm min}$
increases linearly with $d$ up to $d=8\mu$m (similarly to $1/f_{\rm
infl}$ of conductive s-EC), one finds the values of $1/f_{\rm
min}(d)$ close to $\tau_{\rm d}(d)$, and the best fit $1/f_{\rm min}
\propto d^2$ does not differ significantly from $\tau_{\rm d}(d)$
(see inset of Fig.~\ref{Uvfdielddependence}). At $d=20\mu$m $U_{\rm
c}(f)$ becomes a monotonically increasing function (at least above
$f=5$Hz). One has to mention here that in our calculations we have
not resolved the very low frequency range ($\sim 1$Hz), where the
theoretical analysis becomes more difficult due to numerical
problems and the experiments also require special care (and are,
therefore, typically avoided). From Fig. \ref{Uvfdielddependence}
one also sees that the steepness of $U_{\rm c}(f)$ in the high
frequency range (where the function is close to linear) also depends
on $d$; ${\rm d}U_{\rm c}/{\rm d}f$ increases with $d$.

\begin{figure}[h!t]
\includegraphics[width=8.5cm]{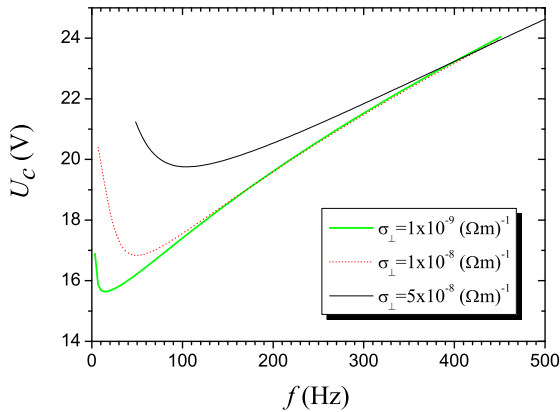}\\
\caption{(Color online) Results of numerical calculations for Phase
5 with $\sigma_a/\sigma_{\perp}=0.1$, $e_1=-26.5$pC/m,
$e_3=-23.6$pC/m and $d=3.4\mu$m for three different conductivities
$\sigma_{\perp}$.} \label{Uvfdielcaniz0_1varycond}
\end{figure}

The other situation is when $\tau_{\rm q}$ is tuned via
$\sigma_{\perp}$ while keeping $d$ and thus $\tau_{\rm d}$ constant
($1/\tau_{\rm d}=75.5$Hz). The influence of $\sigma_{\perp}$ on
$U_{\rm c}(f)$ is illustrated in Fig. \ref{Uvfdielcaniz0_1varycond}.
With the increase of the conductivity (i.e, of $1/\tau_{\rm q})$,
both $f_{\rm min}$ and the lowest threshold value $U_{\rm c}(f_{\rm
min}$) increase.

\begin{figure}[h!t]
\includegraphics[width=8.5cm]{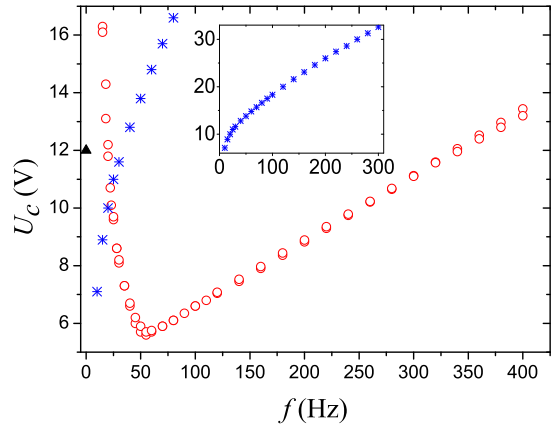}\\
\caption{(Color online) Frequency dependence of the s-EC threshold
voltage $U_{\rm c}$ measured in a sample of Phase 4 with $d=3.2\mu$m
(circles). At dc voltage, the threshold for the flexoelectric
domains is also indicated (solid triangle). For comparison,
experimental data obtained on another sample of Phase 4 with
$d=12.4\mu$m is also plotted (stars). The inset displays
experimental results on the $d=12.4\mu$m cell in the whole frequency
range studied.} \label{Ph4all}
\end{figure}

\begin{figure}[h!t]
\includegraphics[width=8.5cm]{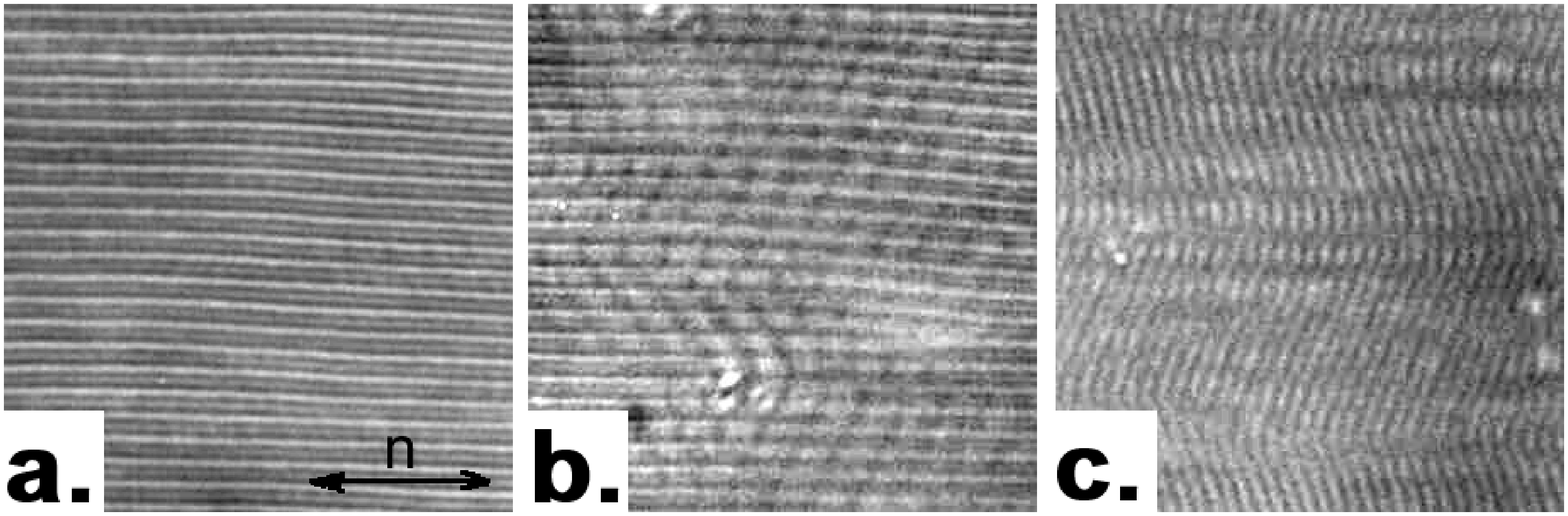}\\
\caption{Snapshots of patterns taken slightly above the threshold in
a $d=3.2\mu$m sample of Phase 4. {\bf (a.)} flexoelectric domains at
dc voltage; {\bf (b.)} coexisting flexoelectric domains and
traveling dielectric oblique rolls at $f=15$Hz; {\bf (c.)} traveling
dielectric oblique rolls at $f=70$Hz. The arrow denotes the initial
director. The physical size of the images is $(65 \times 65) \mu{\rm
m}^2$.} \label{Ph4snapshots1}
\end{figure}

Samples of the nematic Phase 4 have also been investigated at
$T=30^{\circ}$C. Measurements on the frequency dependence of $U_{\rm
c}$ are shown in Fig. \ref{Ph4all} for a $d=3.2\mu$m cell (circles).
At dc voltage (solid triangle) and at low frequencies (up to
$f=15$Hz) flexoelectric domains [see in Fig. \ref{Ph4snapshots1}(a)]
have been observed with a tendency of $U_f$ increasing with the
frequency. At $f=15$Hz the thresholds for flexoelectric domains and
for traveling dielectric oblique rolls almost coincide (see the
coexisting patterns in Fig. \ref{Ph4snapshots1}(b) -- a similar
behavior has also been reported in \cite{May07}). At higher $f$
flexoelectric domains could not be detected. Instead, traveling
dielectric oblique rolls [Fig. \ref{Ph4snapshots1}(c)] or traveling
dielectric normal rolls have been observed. The threshold curve for
these traveling rolls has a minimum similarly to that observed in
dielectric s-EC of Phase 5. In contrast to the thin cell, however,
experimental data obtained on a $d=12.4\mu$m thick sample of Phase 4
(star symbols in Fig. \ref{Ph4all} and in its inset) demonstrate a
monotonically increasing $U_{\rm c}(f)$ (at least for $f \geq
10$Hz). At the same time, Fig. \ref{Ph4all} represents an
experimental support of numerical results shown in Fig.
\ref{Uvfdielddependence} obtained though with Phase 5 parameters
(unfortunately, in Phase 5 samples with $d \geq 10 \mu{\rm m}$,
dielectric s-EC could not be obtained at low enough $f$).

\section{\label{sec:nsEC}Nonstandard EC }

Previous measurements have shown that in samples of {\bf 8/7} with
typical thickness ($d\geq 10\mu$m), ns-EC takes place in the whole
investigated frequency range at $T\leq 85^{\circ}$C, characterized
by a linear $U_{\rm c}(f)$ dependence. In the temperature range
$85^{\circ}$C$<T<90^{\circ}$C both ns-EC and s-EC could be
observed depending on $f$ and $d$, while above $T=90^{\circ}$C
always s-EC has appeared at onset \cite{Toth07}.

\begin{figure}[h!t]
\includegraphics[width=8.5cm]{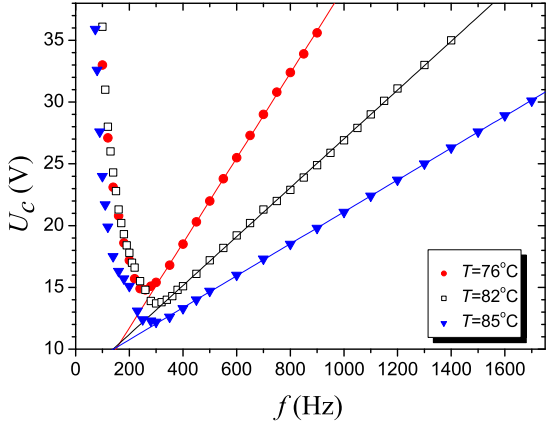}\\
\caption{(Color online) Frequency dependence of the ns-EC threshold
voltage $U_{\rm c}$ measured in a sample of {\bf 8/7} with
$d=3.4\mu$m for three different temperatures $T$. Solid lines
represent linear fit to data at high frequencies (above the minima
of the curves).} \label{8_7Uvsfdifft}
\end{figure}

In accordance with the above experimental findings, in our $d=3.4
\mu$m sample of {\bf 8/7} ns-EC has been found below $T \approx
90^{\circ}$C. At higher frequencies, similarly to thick cells, a
linear frequency dependence of the threshold voltage has been
detected which is illustrated in Fig. \ref{8_7Uvsfdifft} for three
different temperatures. However, at lower $f$ the frequency
dependence of $U_{\rm c}$ is different from the one observed in
thicker samples \cite{Toth07}. The threshold voltage increases
abruptly here when $f \rightarrow 0$ as shown in Fig.
\ref{8_7Uvsfdifft}; i.e., $U_{\rm c}(f)$ becomes a non-monotonic
function with a minimum $U_{\rm cmin}$ at $f_{\rm min}$, similarly
to the case of dielectric s-EC in Phase 5 and Phase 4 samples.

Studies on the temperature dependence of $U_{\rm c}(f)$ in the
sample of {\bf 8/7} with $d=3.4\mu$m have revealed that: {\bf (i.)}
$U_{\rm cmin}$ reduces with the increase of $T$; {\bf (ii.)} $f_{\rm
min}$ increases with $T$; {\bf (iii.)} the steepness of $U_{\rm
c}(f)$ in the high frequency range (where the function is linear) is
also temperature dependent; ${\rm d}U_{\rm c}/{\rm d}f$ diminishes
linearly with the increase of $T$.

\begin{figure}[h!t]
\includegraphics[width=8.5cm]{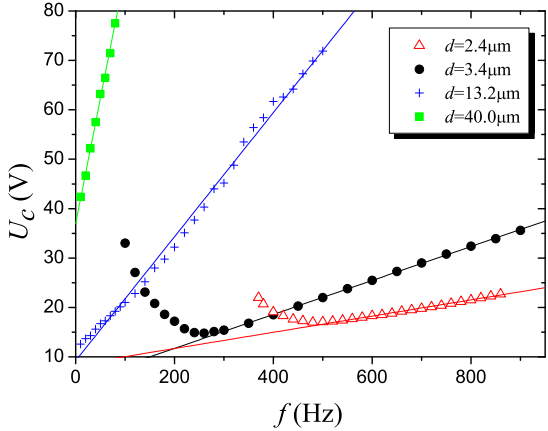}\\
\caption{(Color online) ns-EC threshold curves $U_{\rm c}(f)$
measured in samples of {\bf 8/7} of different thicknesses $d$ at
$T=76^{\circ}$C. Solid lines represent linear fit to data.}
\label{8_7UvfT0_3}
\end{figure}

In an even thinner ($d=2.4\mu$m) sample of {\bf 8/7} no periodic EC
structure could be observed, instead, a large-scale, non-periodic
pattern has appeared at a critical voltage $U_{\rm ls}$. Since a
similar non-periodic, large-scale pattern has also been observed as
a concomitant phenomenon of the ns-EC pattern at $U_{\rm c}$ in
thicker samples, in Fig. \ref{8_7UvfT0_3} we compare this $U_{\rm
ls}$ with $U_{\rm c}$ of ns-EC measured in samples of different
thicknesses at $T=76^{\circ}$C. It is seen that the novel
non-monotonic behavior is restricted to the two smallest thicknesses
($d=2.4 \mu$m and $d=3.4 \mu$m). $f_{\rm min}$ increases as $d$ is
lowered (it is about $220$Hz for the $d=3.4\mu$m sample, and is
roughly $500$Hz for $d=2.4\mu$m). Therefore, one can conclude that
$f_{\rm min}$ depends stronger on $d$, than on the material
parameters (compare with the much weaker temperature dependence of
this minimum in Fig. \ref{8_7Uvsfdifft}). It can be noted that while
$U_{\rm cmin}$ and $f_{\rm min}$ grow, ${\rm d}U_{\rm c}/{\rm d}f$
decreases with lowering the sample thickness $d$.

\begin{figure}[h!t]
\includegraphics[width=8.5cm]{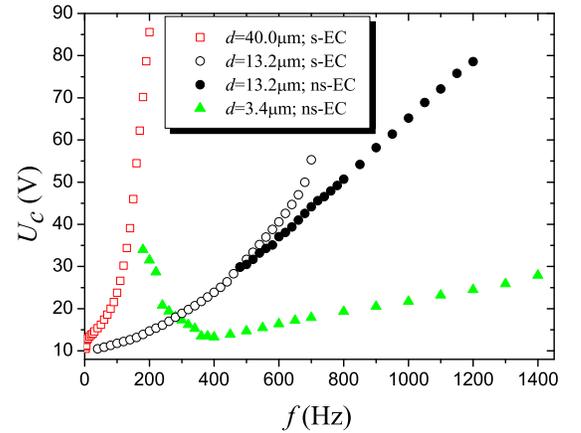}\\
\caption{Frequency dependence of the threshold voltage $U_{\rm c}$
measured in {\bf 8/7} samples of different thickness ($d=3.4\mu$m,
$13.2\mu$m and $40\mu$m) at $T \approx 90^{\circ}$C (in the vicinity
of the s-EC to ns-EC transition). Open symbols denote the s-EC
threshold, while closed symbols stand for ns-EC.}
\label{8_7Uvf0_95ddep}
\end{figure}

The temperature range where ns-EC is observable extends to higher
$T$ in thin samples compared to thicker ones. To demonstrate this,
we plotted in Fig.~\ref{8_7Uvf0_95ddep} the frequency dependence
of the threshold voltage at $T \approx 90^{\circ}$C for three
different sample thicknesses. In the thick, $d=40\mu$m, cell s-EC
exists in the whole accessible frequency range. At an intermediate
thickness of $13.2 \mu$m a s-EC to ns-EC transition has been
observed at $f \approx 500$Hz. In the thin, $d=3.4\mu$m, cell
ns-EC has been detected in the whole frequency range. In this
latter case a non-monotonic $U_{\rm c}(f)$ is observed, just as at
lower temperatures, or in the dielectric s-EC of Phase 5 and Phase
4. Fig.~\ref{8_7Uvf0_95ddep} demonstrates that a s-EC to ns-EC
transition can also be induced by reducing the sample thickness
(while keeping the temperature and frequency fixed), in addition
to the already reported temperature, frequency and electric field
induced transitions \cite{Toth07}.

Recent theoretical advances (the analysis of the extended SM) have
shown that the flexoelectric charge separation mechanism could be
responsible for the occurrence of ns-EC \cite{Krekhov07}. Numerical
simulations based on the extended SM could actually yield a good
agreement with experimental results on a thick sample of the
compound {\bf 8/7} using a reasonable guess for unknown material
parameters. Extension of the numerical analysis to smaller
thicknesses and lower frequencies is, however, not so
straightforward in the parameter range of ns-EC as for s-EC. In
addition some important parameters (e.g., viscosities and
flexoelectric coefficients) should first be determined by
independent measurements. As a consequence a theoretical $U_{\rm
c}(f)$ dependence for Fig.~\ref{8_7Uvf0_95ddep} could not yet been
provided; it remains a target for further studies.

\section{\label{sec:concl}Discussion}

We have reported about measurements of the threshold voltage of
electroconvection patterns in various nematic liquid crystals
focusing mainly on thin ($d\approx 3.1-3.5 \mu m$) samples in which
$1/\tau_{\rm d}$ is relatively large. These samples have exhibited
patterns of various types in the vicinity of $1/\tau_{\rm d}$:
conductive s-EC, dielectric s-EC as well as nonstandard EC. For all
compounds and pattern types we have found a novel frequency
dependence of the threshold voltage at low frequencies, which
deviates significantly from that typically seen in thicker cells
with smaller $1/\tau_{\rm d}$. This indicates that this novel
phenomenon is quite robust. The character of the novel $U_{\rm
c}(f)$ behavior seems, however, to depend on the type of the
pattern. On the one hand, whenever conductive s-EC developes at
onset, $U_{\rm c}(f)$ expresses a pronounced decrease (bending down)
if the frequency is lowered, thus $U_{\rm c}(f)$ has an inflection
point. On the other hand, if the onset pattern is dielectric s-EC in
the vicinity of $1/\tau_{\rm d}$, $U_{\rm c}(f)$ exhibits an abrupt
increase (bending up) towards lower $f$, so $U_{\rm c}(f)$ has a
minimum. In case of ns-EC, the low frequency behavior is similar to
that of the dielectric s-EC. As according to recent theoretical
results the ns-EC is actually a dielectric mode \cite{Krekhov07},
one can conclude that whether $U_{\rm c}(f)$ bends down or up at
$f\rightarrow 0$, depends on the spatiotemporal symmetry of the
solution at onset. The significant changes in the curvature of
$U_{\rm c}(f)$ occur in the frequency range of $1/\tau_{\rm d}$,
even though some characteristic frequencies ($f_{\rm infl}$ for the
conductive or $f_{\rm min}$ for the dielectric pattern) do not scale
precisely with $\tau_{\rm d}$.

Besides the experiments, the threshold voltages have also been
calculated numerically for s-EC in a linear stability analysis
using both the standard model of electroconvection (neglecting
flexoelectricity), and the extended SM (including the
flexoelectric effect). The simulations have shown that in the
absence of flexoelectricity thin as well as thick cells exhibit
the "regular" $U_{\rm c}(f)$ behavior, without the novel
dependence at low $f$. Including, however, flexoelectricity with
properly adjusted flexoelectric coefficients the pronounced low
frequency decrease of the conductive s-EC threshold could be
reproduced in a quantitative agreement with the experimental data.

Unfortunately, a similar quantitative match could not be achieved
for dielectric s-EC patterns. Nevertheless, simulation experiments
have shown that even the abrupt increase of the dielectric $U_{\rm
c}(f)$ in thin cells at low $f$ can be qualitatively reproduced by
the extended SM if one uses (together with the same flexoelectric
coefficients as for the conductive s-EC) a $\sigma_a/\sigma_{\bot}$
value which is lower than expected from the measurements. One has to
emphasize here that the fitted $e_1$ and $e_3$ parameters fall into
the range of values measured for nematics, however, we do not claim
that the fitted values are unique, some other $e_1$, $e_3$
combinations might provide almost as good match with experimental
data. Furthermore, in the numerical calculations only a subset of
parameters ($\sigma_a$, $\sigma_{\perp}$, $e_1$, $e_3$) have been
varied. One can not exclude that by slight adjustment of the other
parameters (allowed by experimental errors in measuring the values
in Table \ref{table1}) a better agreement between measurements and
calculations could be obtained.

The numerical simulations presented above have confirmed that the
experimentally detected novel low frequency dependencies of $U_{\rm
c}(f)$ are consequences of the flexoelectricity and represent a
competition between flexoelectric and Coulomb charge separation
mechanisms. The frequency range in which it occurs is in the order
of $1/\tau_{\rm d}$, thus it can be tuned by the proper choice of
the sample thickness $d$.

In the paper we have focused on the EC threshold voltages only. For
a complete analysis of the onset behavior one should also confront
the experimental and theoretical data regarding the critical
wave-vector ${\bf q}_c$ of the pattern. The special experimental
conditions, namely the small wavelength of the pattern (about the
same -- few microns -- for conductive and dielectric s-EC as well as
for ns-EC), the fast traveling speed of the rolls, the relatively
weak contrast (weaker than in thick samples) unfortunately did not
allow to obtain ${\bf q}_c$ values with required precision by
polarizing microscopy.

From the misfit of the dielectric thresholds one can draw a double
conclusion. On the one hand, the obtained qualitative agreement may
indicate that the extended SM does contain the main ingredients of
the physical mechanism responsible for the formation of dielectric
patterns even under our experimental conditions. On the other hand,
however, the lack of quantitative match implies that the
applicability limits of the model could be approached by testing
thin cells at low frequencies and inclusion of additional phenomena
(resulting in an apparent reduction of $\sigma_a/\sigma_{\bot}$)
would be needed for a complete description. One such candidate could
be the (unipolar) charge injection at the electrodes which is not
included in the SM. For a low value of the overall conductivity (as
in our case for dielectric s-EC) the weak charge focusing effect may
be disrupted by charge injection at dc or at low frequency ac fields
as it has been suggested recently \cite{Krishnamurthy07}.

We also have to note here that in our thin samples for all compounds
traveling EC rolls have been detected at the onset. It has been
proven that traveling waves can be explained if the assumption of
Ohmic conductivity is given up and the SM is replaced by the weak
electrolyte model (WEM) \cite{Treiber95} taking into account ionic
phenomena. It has also been shown that for typical EC samples ($d
\gtrsim 10 \mu$m) in the conductive s-EC regime the $U_{\rm c}$ and
${\bf q}_c$ values provided by the WEM differ by less than $1\%$
from those calculated with the SM \cite{Treiber97}. However, in this
comparison flexoelectricity has not been included. Neither has the
test been performed for smaller $d$ values, nor for dielectric s-EC.
Therefore, a more complete description of the observed phenomena
would need the inclusion of the flexoelectric effect into the WEM,
and an extension of WEM based numerical studies to dielectric s-EC
as well as to ns-EC, which is going far beyond the scope of the
present paper. Such simulations would require some knowledge about
additional material parameters, such as ionic mobility,
recombination rate, equilibrium ion density. In this respect the
measurements of the dc and low frequency ac conductivities of thin
nematic layers could be important to clarify the impact of weak
electrolyte effects. Future studies in these directions could
possibly bridge the quantitative gap between the experiments and
numerical data presented in this paper for the dielectric s-EC.

\begin{acknowledgments}
The authors thank to W. Pesch for fruitful discussions and for
providing the numerical code. The authors are grateful to G. Pelzl
for providing the liquid crystal \textbf{8/7}. Financial support by
the Hungarian Research Fund OTKA-K61075, DFG Grant No. Kr690/22-1
and SFB 481/A8 are gratefully acknowledged.
\end{acknowledgments}


\begin{thebibliography}{15}
\bibitem{remark} Here, we kept the notations for different modes as introduced in
Ref. \cite{Krekhov07}, where the full description of the model is
given. In Ref. \cite{Zimmer91} notations of IA, IIA, IB and IIB have
been used for modes I, II, III and IV, respectively.
\end{thebibliography}
\end{document}